\documentclass[12pt]{article}
\usepackage{latexsym}
\usepackage{amsmath, amsfonts, amsthm,hyperref,url}
\usepackage{xcolor,tikz}
\usepackage{graphics}
\usepackage{epsfig}
\usepackage{stmaryrd}
\usepackage{multicol}
\usepackage{pdfsync}
\usepackage{dsfont}
\usepackage{comment}
\usepackage{algorithmic}
\usepackage{algorithm}
\usepackage{caption}
\usepackage{psfrag}

\usepackage{textcomp}

\theoremstyle{definition}

\theoremstyle{remark}

\numberwithin{equation}{section}

\usetikzlibrary{fit,positioning,arrows.meta}
\tikzset{neuron/.style={shape=circle, minimum size=1.25cm, 
  inner sep=0, draw, font=\small}, io/.style={neuron, fill=gray!20}}
  
\usepackage{setspace}
\title{{\large\bf Universal features of price formation in financial markets:\\
perspectives from Deep Learning} }
\date{\large \today}
\author{ {\large Justin Sirignano}\footnote{University of Illinois at Urbana-Champaign.} \phantom{.}  {\large and Rama Cont}\footnote{CNRS \& Department of Mathematics, Imperial College London. Corresponding author.}  \thanks{The authors thank seminar participants at the London Quant Summit 2018, JP Morgan and Princeton University for their comments. Computations for this paper were performed using a grant from the CFM-Imperial Institute of Quantitative Finance and the Blue Waters supercomputer grant ``Distributed Learning with Neural Networks". }
}

\begin{document}
\maketitle

%supply and demand
%nonlinear relationship between order book and price move, supply and demand and price dynamics
%liquidity
%ask and bid levels
%stationarity

\begin{abstract}
Using a  large-scale Deep Learning approach applied to a high-frequency database containing billions of electronic market quotes and transactions for US equities, we uncover nonparametric evidence for the existence of a {\it universal} and {\it stationary} price formation mechanism relating the dynamics of supply and demand for a stock, as revealed through the order book, to subsequent variations in its market price.
We assess the model by testing its out-of-sample predictions for the direction of price moves given the history of price and order flow, across  a wide range of stocks and time periods.
The  universal price formation model exhibits a remarkably stable out-of-sample prediction accuracy across  time, for a wide range of stocks from different sectors. Interestingly, these results also hold for stocks which are not part of the training sample, showing that the relations captured by the model are universal and not asset-specific.

The universal model --- trained on data from all stocks --- outperforms, in terms of out-of-sample prediction accuracy,  asset-specific  linear and nonlinear  models trained on time series of any given stock, showing that the universal nature of price formation weighs in favour of pooling together financial data from various stocks, rather than designing asset- or sector-specific models as  commonly done. 
Standard data normalizations based on volatility, price level or average spread, or partitioning the training data into sectors or categories such as large/small tick stocks, do not improve training results. On the other hand, inclusion of price and order flow history over many past observations improves forecasting performance, showing evidence of path-dependence  in price dynamics.

\end{abstract}
\newpage
\tableofcontents
\section{Price formation: how market prices react to supply and demand}

The computerization of financial markets and the  availability of detailed electronic records of order flow and price dynamics in financial markets over the last  decade has unleashed TeraBytes of high frequency data on  transactions, order flow and order book dynamics  in listed markets, which provide us with a detailed view of the high-frequency dynamics of supply, demand and price  in these markets  \cite{IEEE}.
This data may be put to use to explore the nature of the price formation mechanism which describes  how market prices react to fluctuations in supply and demand.
At a high level, a `price formation mechanism' is a map which represents the correspondence between the market price and variables such as price history and order flow: 

\begin{center} Price($t+\Delta t$) = F\big{(}Price history(0...t), Order Flow(0...t), Other Information\big{)}= $F(X_t,\epsilon_t)$, \end{center} \notag
where $X_t$ is a set of state variables (e.g., lagged values of price, volatility, and order flow), endowed with some dynamics and $\epsilon_t$ is a random `noise' or innovation term representing the arrival of new information and other effects not captured entirely by the state variables.
Empirical and theoretical market microstructure models, stochastic models and machine learning price prediction models can all be viewed as different ways of representing  this map $F$, at various time resolutions $\Delta t$.

One question, which has been implicit in the literature, is the degree to which this map $F$ is {\it universal} (i.e., independent of the specific asset being considered). The generic, as opposed to asset-specific, formulation of market microstructure models seems to implicitly assume such a universality. Empirical evidence on the universality of certain stylized facts \cite{empirical2001} and scaling  relations \cite{DonierBouchaud,Bondarenko2017,KyleObizhaeva,Mandelbrot1997} seems to support the universality hypothesis. Yet, the practice of statistical modeling of financial time series has remained asset specific: when building a model for the returns of a given asset, market practitioners and econometricians only use data from the same asset. For example, a model for Microsoft would only be estimated using Microsoft data, and would not use data from other stocks. 

Furthermore, the data used for estimation is often limited to a recent time window, reflecting the belief that financial data can be 
`non-stationary' and prone to regime changes which may render older data less relevant for prediction.

Due to such considerations, models considered in financial econometrics, trading and risk management applications are asset-specific and their parameters are (re)estimated over time using a time window of recent data. That is, for  asset $i$ at time $t$ the model assumes the form
\begin{equation} \textrm{Price}_i(t+\Delta t)= F\big{(} X^i_{ 0: t},  \epsilon_t\  | \phantom{..} \theta_{i}(t)\big{)}, \label{eq.F}  \notag \end{equation} where the model parameter $\theta_{i}(t)$ is periodically updated using recent data on  price and other state variables related to asset $i$. As a result, data sets are fragmented across assets and time and, even in the high frequency realm, the size of data sets used for model estimation and training are  orders of magnitude smaller than those encountered in other fields where Big Data analytics have been successfully applied. This is one of the reasons why, except in a  few instances  \cite{buhler,DixonSequenceClassification,JPM,SirignanoDeepLearningLimitOrderBook,DeepLearningMortgages}, large-scale learning methods such as Deep Learning \cite{GoodFellow} have not been deployed for quantitative modeling in finance. In particular, the non-stationarity argument is sometimes invoked to warn against their use.

On the other hand, if the relation between these variables were {\it universal} and {\it stationary}, i.e. if the parameter $\theta_i(t)$ varies {\it neither} with the asset $i$ {\it nor} with time $t$, then one could potentially pool data across different assets and time periods and use a much richer data set to estimate/ train the  model. For instance, data on a flash crash episode in one asset market could provide insights into how the price of another asset would react to severe imbalances in order flow, whether or not such an episode has occurred in its history.

In this work, we provide evidence for the existence of such a universal, stationary relation between order flow and market price fluctuations, using a nonparametric approach based on Deep Learning. Deep learning can estimate nonlinear relations between variables using `deep' multilayer neural networks which are trained on large data sets using `supervised learning' methods \cite{GoodFellow}.

Using a  deep neural network architecture trained on a  high-frequency database containing billions of electronic market transactions and quotes for US equities, we uncover nonparametric evidence for the existence of a {\it universal} and {\it stationary} price formation mechanism relating the dynamics of supply and demand for a stock, as revealed through the order book, to subsequent variations in its market price.
We assess the model by testing its out-of-sample predictions for the direction of price moves given the history of price and order flow, across  a wide range of stocks and time periods.
The  universal price formation model exhibits a remarkably stable out-of-sample prediction accuracy across  time, for a wide range of stocks from different sectors. Interestingly, these results also hold for stocks which are not part of the training sample, showing that the relations captured by the model are universal and not asset-specific.
We  observe that the neural network thus trained  outperforms linear models,  pointing to  the presence of nonlinear  relationships between order flow and price changes.

Our paper provides quantitative evidence for the existence of a universal  price formation mechanism in financial markets. 
The universal nature of the price formation mechanism is reflected by the fact that a model trained on data from all stocks outperforms, in terms of out-of-sample prediction accuracy,  stock-specific  linear and nonlinear  models trained on time series of any given stock. This shows that the universal nature of price formation weighs in favour of pooling together financial data from various stocks, rather than designing stock- or sector-specific models as  commonly done. 
Also, we observe that standard data transformations such as normalizations based on volatility or average spread, or partitioning the training data into sectors or categories such as large/small tick stocks, do not improve training results. On the other hand, inclusion of price and order flow history over many past observations improves forecasting performance, showing evidence of path-dependence in price dynamics.

Remarkably, the universal model is able to extrapolate, or  \emph{generalize}, to  stocks  not within the training set. The universal model is able to perform well on completely new stocks whose historical data the model was never trained on. This implies that the universal model captures features of the price formation mechanism  which are robust across stocks and sectors. This feature alone is quite interesting for applications in finance where missing data problems and newly issued securities often complicate model estimation.

{\bf Outline} %The outline of the article is as follows.
Section \ref{sec.DeepLearning} describes the dataset and the supervised learning approach used to extract information about the price formation mechanism.
Section \ref{sec.Results} provides evidence for the existence of a universal and stationary relationship  linking order flow and price history to price variations.
Section \ref{sec.conclusion} summarizes our main findings and discusses some implications.

%\section{Data and Computations} \label{sec.Dataset}

\section{A data-driven model of  price formation via Deep Learning} \label{sec.DeepLearning}

Applications such as image, text, and speech recognition have been revolutionized by the advent of `Deep Learning' --- the use of multilayer (`deep') neural networks trained on large data sets  to uncover complex nonlinear relations between high-dimensional inputs (`features') and outputs \cite{GoodFellow}. 

At an abstract level, a deep neural network represents a functional relation $y=f(x)$ between a high-dimensional input vector $x$ and an output $y$ through iterations (`layers') consisting of weighted sums followed by the application of nonlinear `activation' functions. Each iteration corresponds to a `hidden layer' and a deep neural network can have many hidden layers. Neural networks can be used as `universal approximators' for complex nonlinear relationships \cite{Hornik}, by appropriately choosing the weights in each layer. 

In supervised learning approaches, network weights are estimated by optimizing a regularized cost function reflecting in-sample discrepancy between the network output and desired outputs. In a deep neural network, this represents a high-dimensional optimization over hundreds of thousands (or even millions) of parameters. This optimization is computationally intensive due to the large number of parameters and large amount of data. Stochastic gradient descent algorithms (e.g., RMSprop or ADAM) are used for training neural networks, and training is parallelized on Graphics Processing Units (GPUs).

We apply this approach to learn the relation between supply and demand on an electronic exchange --- captured in the history of the order book for each stock --- and the subsequent variation of the market price. Our data set is a high-frequency record of  all orders, transactions and order cancellations for approximately 1000 stocks traded on the NASDAQ between January 1, 2014 and  March 31, 2017.\footnote{Historical order book data was reconstructed from NASDAQ Level III data using the LOBSTER data engine \cite{Huang}.}  
\begin{figure}[h!]
\centering
\includegraphics[scale=0.40]{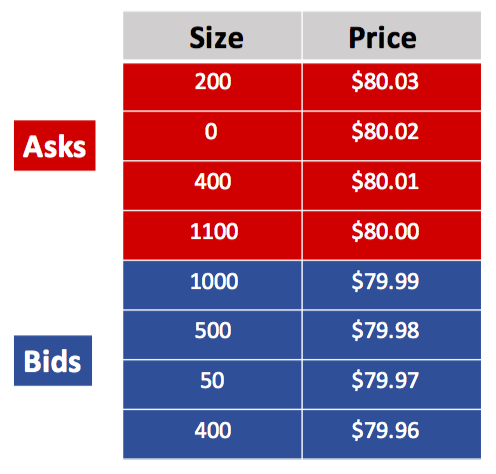}
\caption{The limit order book represents a snapshot of the supply and demand for a stock on an electronic exchange. The `ask' side represents sell orders and  the `bid' side, buy orders.  The size represents the number of shares available for sale/purchase at a given price.  The difference between the lowest sell price (ask) and the highest buy price (bid) is the `spread' (in this example, $1$ \textcent ).}
\label{LOBexampleFigure}
\end{figure}

 Electronic buy and sell orders are continuously submitted, cancelled and executed through the exchange's \emph{order book}.   A `limit order' is a buy or sell order for a stock at a certain price and will appear in the order book at that price and remain there until cancelled or executed. The `limit order book' is a snapshot of all outstanding limit orders and thus represents the visible supply and demand for the stock (see Figure \ref{LOBexampleFigure}). In US stock markets, orders can be submitted at prices occurring at multiples of $1$ cent.  The `best ask price' is the lowest sell order and the `best bid price' is the highest bid price. The best ask price and best bid price are the prices at which the stock can be immediately bought or sold. The `mid-price' is the average of the best ask price and best bid price. The order book evolves over time as new orders are submitted, existing orders are cancelled, and trades are executed.

In electronic markets such as the NASDAQ, new orders may arrive at high frequency --- sometimes every microsecond ---  and order books of certain stocks can update millions of times per day. This leads to TeraBytes of data, which we put to use to build a {\bf data-driven model} of the price formation process.

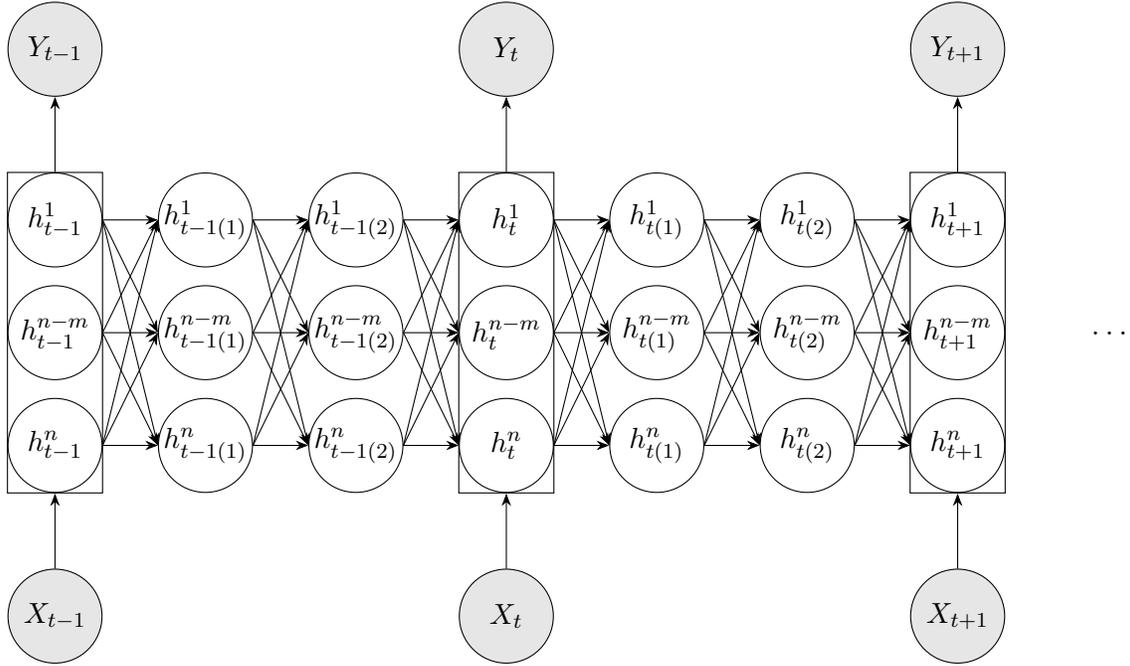
\begin{figure} 
\center
\begin{tikzpicture}[x=2cm, y=1.5cm, >=Stealth]
\foreach \jlabel [count=\j, evaluate={\k=int(mod(\j-1,3)); \jj=int(\j-1);}]
  in {t-1, t-1(1), t-1(2), t, t(1), t(2), t+1}{
    \foreach \ilabel [count=\i] in {1, n-m, n}
        \node [neuron] at (\j, 1-\i) (h-\i-\j){$h_{\jlabel}^{\ilabel}$};
    \ifcase\k
      \node [fit=(h-1-\j) (h-3-\j), inner sep=0, draw] (b-\j) {};
      \node [io, above=of b-\j] (y-\j) {$Y_{\jlabel}$};
      \node [io, below=of b-\j] (v-\j) {$X_{\jlabel}$};
      \draw [->] (v-\j) -- (b-\j);
      \draw [->] (b-\j) -- (y-\j);
    \fi
    \ifnum\j>1
      \foreach\i in {1, 2, 3}
        \foreach \ii in {1, 2, 3}
           \draw [->] (h-\i-\jj.east) -- (h-\ii-\j.west);
    \fi
} 
%\node [left=of h-2-1] {\ldots};
%
\node [right=of h-2-7] {\ldots};
\end{tikzpicture}
\caption{Architecture of a recurrent neural network.}
\label{fig.LSTM}
\end{figure}

When the input data is a time series, causality constraints require that the relation between input and output respects the ordering in time. Only the past may affect the present.  A network architecture which reflects this constraint is a recurrent network (see an example in Figure \ref{fig.LSTM}) based on Long Short-Term Memory (LSTM) units \cite{LSTM}.

Each LSTM unit has an internal state which maintains a nonlinear representation of all past data. This internal state is updated as new data arrives. Our network  has $3$ layers of LSTM units followed by a final feed-forward layer of rectified linear units (ReLUs). A probability distribution for the next price move is produced by applying a softmax activation function. LSTM units are specially designed to efficiently encode the temporal sequence of data \cite{LSTM,GoodFellow}.

%\begin{figure} 
%\begin{tikzpicture}[x=2cm, y=1.5cm, >=Stealth]
%\foreach \jlabel [count=\j, evaluate={\k=int(mod(\j-1,3)); \jj=int(\j-1);}]
%  in {t-1, t-1(1), t-1(2), t, t(1), t(2), t+1}{
%    \foreach \ilabel [count=\i] in {1, n-m, n}
%        \node [neuron] at (\j, 1-\i) (h-\i-\j){$h_{\jlabel}^{\ilabel}$};
%    \ifcase\k
%      \node [fit=(h-1-\j) (h-3-\j), inner sep=0, draw] (b-\j) {};
%      \node [io, above=of b-\j] (y-\j) {$Y_{\jlabel}$};
%      \node [io, below=of b-\j] (v-\j) {$X_{\jlabel}$};
%      \draw [->] (v-\j) -- (b-\j);
%      \draw [->] (b-\j) -- (y-\j);
%    \fi
%    \ifnum\j>1
%      \foreach\i in {1, 2, 3}
%        \foreach \ii in {1, 2, 3}
%           \draw [->] (h-\i-\jj.east) -- (h-\ii-\j.west);
%    \fi
%} 
%%\node [left=of h-2-1] {\ldots};
%%
%\node [right=of h-2-7] {\ldots};
%\end{tikzpicture}
%\caption{Architecture of a recurrent neural network.}
%\label{fig.LSTM}
%\end{figure}

We train the network to forecast the next price move from a vector of state variables, which encode the history of the order book over many observation lags. The index $t$ represents the number of price changes. At a high level, the LSTM network is of the form
\begin{eqnarray}
(Y_t, h_t) = f( X_t, h_{t-1}; \theta).
\end{eqnarray}
$Y_t$ is the prediction for the next price move, $X_t$ is the state of the order book  at time $t$,  $h_t$ is the internal state of the deep learning model, reprenting information extracted from the {\it history} of $X$ up to $t$, and $\theta$ designates the model parameters, which correspond to the weights in the neural network. At each time point $t$ the model uses  the current value of state variables $X_t$ (i.e. the current order book) and the nonlinear representation of all previous data $h_{t-1}$, which summarizes relevant features of the history of order flow, to predict the next price move. In principle, this allows for arbitrary history-dependence: the history of the state variables $(X_{s}, s\leq t)$ may affect the evolution of the system, in particular price dynamics, at all future times $T\geq t$ in a nonlinear way. Alternative modeling approaches typically do not allow the flexibility of blending nonlinearity and history-dependence in this manner.

% which loss function??
A supervised learning approach is used to learn the value of the (high-dimensional) parameter $\theta$ by minimizing a regularized negative log-likelihood objective function using a stochastic gradient descent algorithm \cite{GoodFellow}. The parameter $\theta$ is assumed to be constant across time, so it affects the output at all times in a recursive manner. A stochastic gradient descent step at time $t$ requires calculating the sensitivity of the output to $\theta$, via a chain rule, back through the previous times $t-1, t-2, \ldots, t-T$  (commonly referred to as `backpropagation through time'). In theory, backpropagation should occur back to time $0$ (i.e., $T = t$). However, this is computationally impractical and we truncate the backpropagation at some lag $T$. In Section \ref{sec.temporal}, we discuss the impact of the past history of the order book and the `long memory' of the market.

The resulting LSTM  network involves up to hundreds of thousands of parameters. This is relatively small compared to networks used for instance in image or speech recognition, but it is huge compared to econometric models traditionally used in finance. Previous literature has been almost entirely devoted to linear models or stochastic models with a very small number of parameters. It is commonly believed that financial data is far too noisy to build such large models without overfitting; our results show that this is not necessarily true.

Given the size of the data and the large number of network parameters to be learned,
significant computational resources are required both for pre-processing the data and training the network. Training of deep neural networks can be highly parallelized on GPUs. Each GPU has thousands of cores, and training is typically $10 \times$ faster on a GPU than a standard CPU.  The NASDAQ data was  filtered to create training and test sets. This data processing is parallelized over approximately $500$ compute nodes. Training of asset-specific models was also parallelized, with each stock assigned to a single GPU node. Approximately $500$ GPU nodes are used to train the stock-specific models. 

These asset-specific models, trained on the data related to a single stock, were then compared to a `universal model'  trained on the combined data from all the stocks in the dataset. Data from various stocks were pooled together for this purposes without any specific normalization.

\begin{figure}[ht!]
\centering
\includegraphics[scale=0.5]{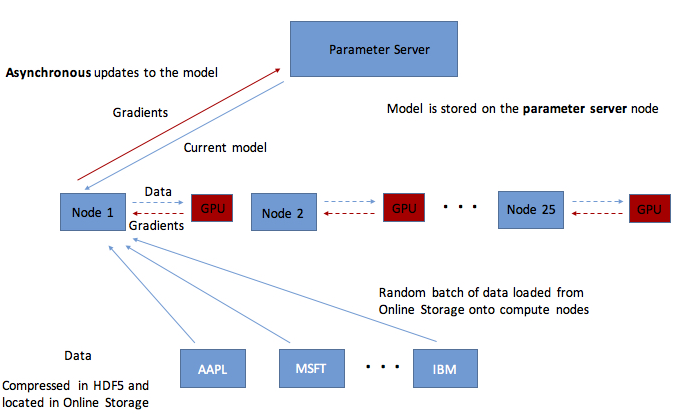}
\caption{Asynchronous stochastic gradient descent for training the neural network.  The dataset, which is too large to be held in the nodes' memories, is stored on the Online Storage system.  Batches of data are randomly selected from all stocks and sent to the GPU nodes.  Gradients are calculated on the GPUs and then the model is asynchronously updated.}
\label{ASGD}
\end{figure}

 Due to the large amount of data, we distributed the training of the universal model across $25$ GPU nodes using asynchronous stochastic gradient descent (Figure \ref{ASGD}). Each node loads a batch of data (selected at random from all stocks in the dataset), computes gradients of the model on the GPU, and then updates the model. Updates occur \emph{asynchronously}, meaning node $j$ updates the model without waiting for nodes $i \neq j$ to finish their computations.  

\section{Results} \label{sec.Results}

We split the universe of stocks into two groups of roughly 500 stocks; training is done on transactions and quotes for stocks from the first group.
We distinguish:
\begin{itemize}
\item stock-specific models, trained using data on all transactions and quotes for a specific stock.
\item the `universal model', trained using data on all transactions and quotes for \textbf{all stocks} in the training set.
\end{itemize}
All models are trained for predicting the direction of the \emph{next price move}. Specifically, if $\tau_1, \tau_2, \ldots$ are the times at which the mid-price $P_t$ changes, we estimate $\mathbb{P} [ P_{\tau_{k+1}} -P_{\tau_{k}} > 0 | X_{\tau_{0:k}} ]$ and $\mathbb{P} [ P_{\tau_{k+1}} -P_{\tau_{k}} < 0 | X_{\tau_{0:k}} ]$  where $X_t$ is the state of the limit order book at time $t$. The models therefore predict whether the next price move is up or down. The  events are irregularly spaced in time. The time interval $\tau_{k+1} - \tau_k$ between price moves can vary considerably from a fraction of a second to seconds. 

We measure the  forecast accuracy of a model for a given stock via the proportion of observations for which it correctly predicts the direction of the next price move. This can be estimated using the empirical estimator
\begin{eqnarray}
A_i= \frac{\textrm{Number of price changes where model correctly predicts price direction for stock $i$}}{\textrm{Total number of price changes} } \times 100\%.  \notag
\end{eqnarray}
All results are \emph{out-of-sample in time}. That is, the accuracy is evaluated on time periods outside of the training set. Model accuracy is reported via the cross-sectional distribution of the accuracy score $A_i$ across stocks in the testing sample, and models are compared by comparing their accuracy scores.

In addition, we evaluate the accuracy of the universal model for stocks {\it outside} the training set. Importantly, this means we assess forecast accuracy for stock $i$ using a model which is trained {\it without} any data on stock $i$. This tests whether the universal model can generalize to completely new stocks.

Typically, the out-of-sample dataset is a 3-month time period. In the context of high-frequency data, 3 months corresponds to millions of observations and therefore provides a lot of scope for testing model performance and estimating model accuracy. In a data set with no stationary trend (as in the case at such high frequencies), a random forecast (`coin-flip') would yield an expected score of $ 50\%$. Given the large size of the data set, even a small deviation (i.e. 1\%) from this 50\% benchmark is statistically significant.

%We now describe the results obtained using . 
The main findings of our data-driven approach may be summarized as follows:
\begin{itemize}
\item {\bf Nonlinearity}: Data-driven models trained using deep learning substantially outperform linear models in terms of forecasting accuracy  (Section \ref{DeepLearningVersusLinear}).
\item {\bf Universality}: The model uncovers universal features that are common across all stocks
(Section \ref{UniversalModelResults}). These features generalize well: they are also observed to hold for stocks which are not part of the training sample. 
\item {\bf Stationarity}: model performance in terms of price forecasting accuracy is remarkably stable across time, even a year out of sample. This shows evidence for the existence of a {\it stationary} relationship between order flow and price changes (Section \ref{sec.stationarity}), which is stable over long time periods. 
\item {\bf  Path-dependence and long-range dependence}: inclusion of price and order flow history is shown to substantially increase the forecast accuracy. This provides evidence that price dynamics depend not only on the current or recent state of the limit order book but on its {\it history}, possibly over  long time scales (Section \ref{sec.temporal}).
\end{itemize}

%Section \ref{SupplyAndDemand} analyzes the relationship between price dynamics and supply and demand in the order book.

Our results show that there is far more common structure across data from different financial instruments than previously thought. Providing a suitably flexible model is used which allows for nonlinearity and history-dependence, data from various assets may be pooled together to yield a data set  large enough  for deep learning.

\subsection{Deep Learning versus Linear Models} \label{DeepLearningVersusLinear}

Linear state space models,  such as Vector Autoregressive (VAR) models, have been widely used in the modeling of high frequency data and in  empirical market microstructure research \cite{Hasbrouck} and provide a natural benchmark for evaluating the performance of a forecast.
Linear models are easy to estimate and  capture in a simple way the trends, linear correlations and autocorrelations in the state variables.

\begin{figure}[h!]
\begin{center}
\includegraphics[width=.4\textwidth, height=50mm]{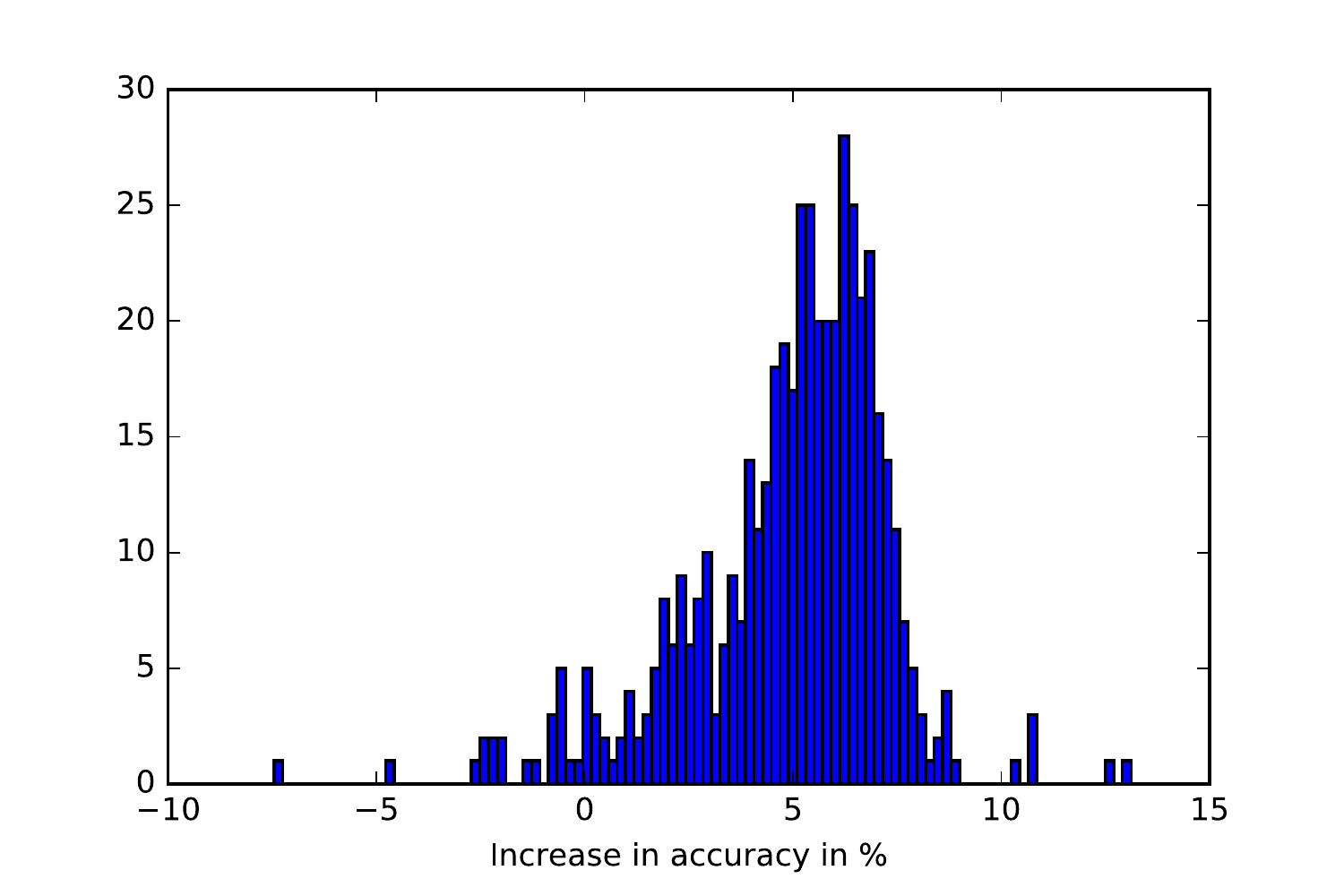}
\includegraphics[width=.4\textwidth, height=50mm]{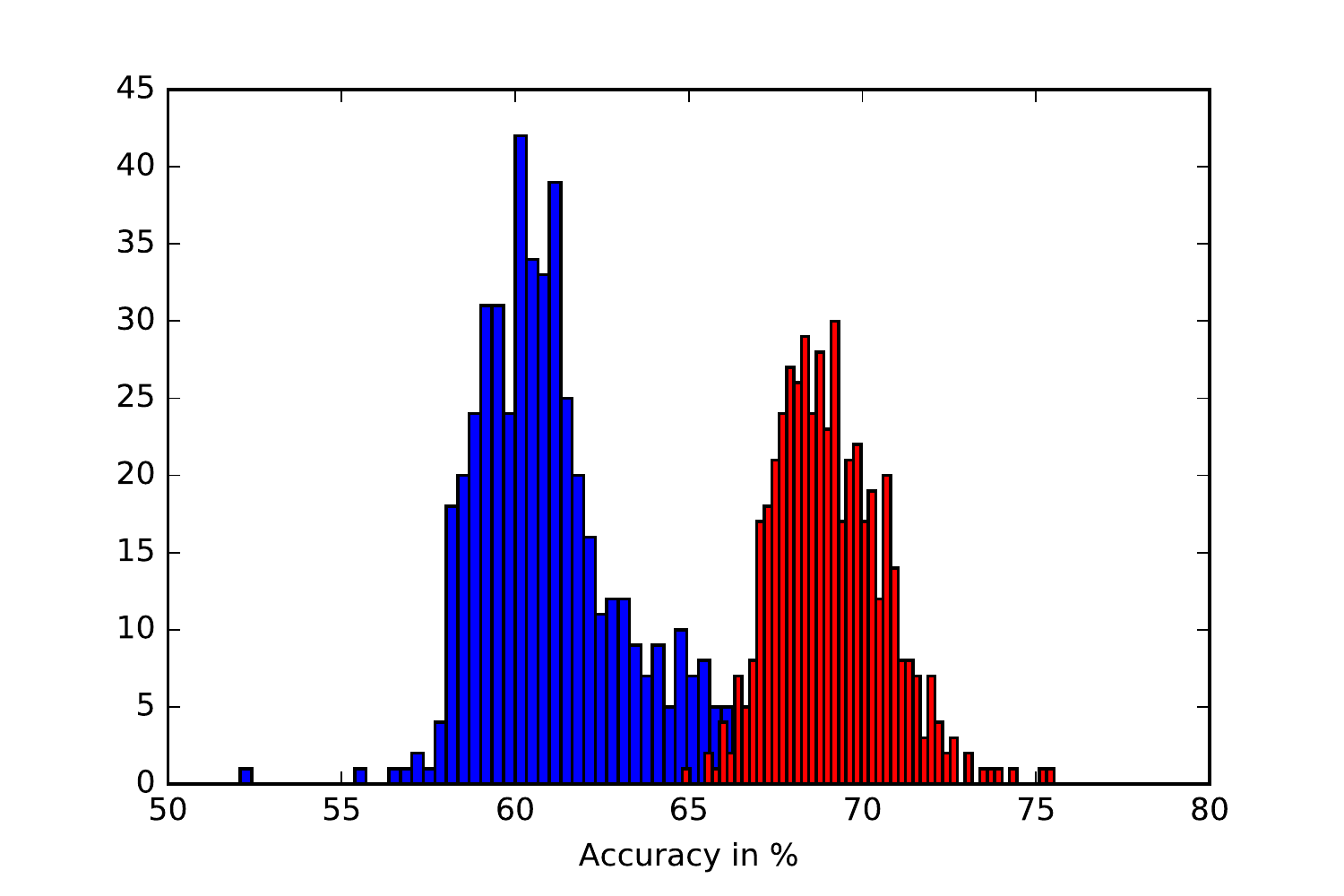}
\end{center}
\caption{Comparison of a deep neural network with linear models. Models are trained to predict the direction $\{-1, +1\}$ of next mid-price move. Comparison for approximately $500$ stocks and out-of-sample results reported for June-August, 2015. Left-hand figure: increase in accuracy of stock-specific deep neural networks versus stock-specific linear models. Right-hand figure: accuracy of a universal deep neural network (\textcolor{red}{red}) compared to stock-specific linear models (\textcolor{blue}{blue}).}
\label{SubplotFigureContour}
\end{figure}

 The results in Figure \ref{SubplotFigureContour} show that the deep learning models substantially outperform linear models. Given the large sample size, an increase of $1\%$ in accuracy is considered significant in the context of high-frequency modeling. 

The linear (VAR) model may be formulated as follows: at each observation we update a vector of {\it linear} features $h_t$ and then use a probit model for the conditional probability of an upward  price move given the state variables:
\begin{eqnarray}
h_t &=& A h_{t-1} + B X_t, \notag \\
Y_t &=& \mathbb{P}( \Delta P_t>0 | X_t, h_t )= G( C X_t + D h_t ).
\end{eqnarray}
where $G$ depends on the distributional assumptions on the innovations in the linear model.
For example, if we use a logistic distribution for the innovations in the linear model, then the probability distribution of the next price move is given by softmax (logistic) function applied to a linear function of  the current order book and linear features: 
$$ \mathbb{P}( \Delta P_t>0 | X_t, h_t )=\textrm{Softmax}( C X_t + D h_t ).$$

We  compare the neural network against a linear model for approximately $500$ stocks.
To compare models we report the difference in accuracy scores across the same test data set. 
Let \begin{itemize}
 \item $L_i$ be the accuracy of the stock-specific linear model $g_{\theta_i}$ for asset $i$ estimated on data only from stock $i$, 
 \item $\hat A_i$ be the accuracy of the stock-specific deep learning model $f_{\theta_i}$ trained on data only from stock $i$, and
 \item $A_i$ be the accuracy for asset $i$ of the {\it universal } deep learning model $f_{\theta}$  trained on a pooled data set of all quotes and transactions for all stocks.
 \end{itemize} 
 The left plot in Figure \ref{SubplotFigureContour} reports the cross-sectional distribution for the increase in accuracy $ \hat A_i - L_i$ when moving from the stock-specific linear model to the stock-specific deep learning model. We observe a substantial increase in accuracy, between 5\% to 10\% for most stocks, when incorporating nonlinear effects using the neural networks. 
 
 The right plot in Figure \ref{SubplotFigureContour} displays histograms of $A_i$ (red) and $L_i$ (blue). We clearly observe that moving from a stock-specific linear model to the universal nonlinear model trained on all stocks substantially improves the forecasting accuracy by around 10\%.

The deep neural network outperforms the linear model since it is able to estimate nonlinear relationships between the price dynamics and the order book, which represents the visible supply and demand for the stock. This is consistent with an abundant empirical and econometric literature documenting nonlinear effects in financial time series, but the large amplitude of this improvement can be attributed to the flexibility of the neural network in representing nonlinearities. 

%Unlike parametric models, we are not constraining the functional form of the nonlinearity.

More specifically, sensitivity analysis of our data-driven model uncovers stable nonlinear relations between state variables and price moves, i.e. {\it nonlinear features} which are useful for forecasting.
\begin{figure}[ht!]
\begin{center}
\includegraphics[width=.47\textwidth]{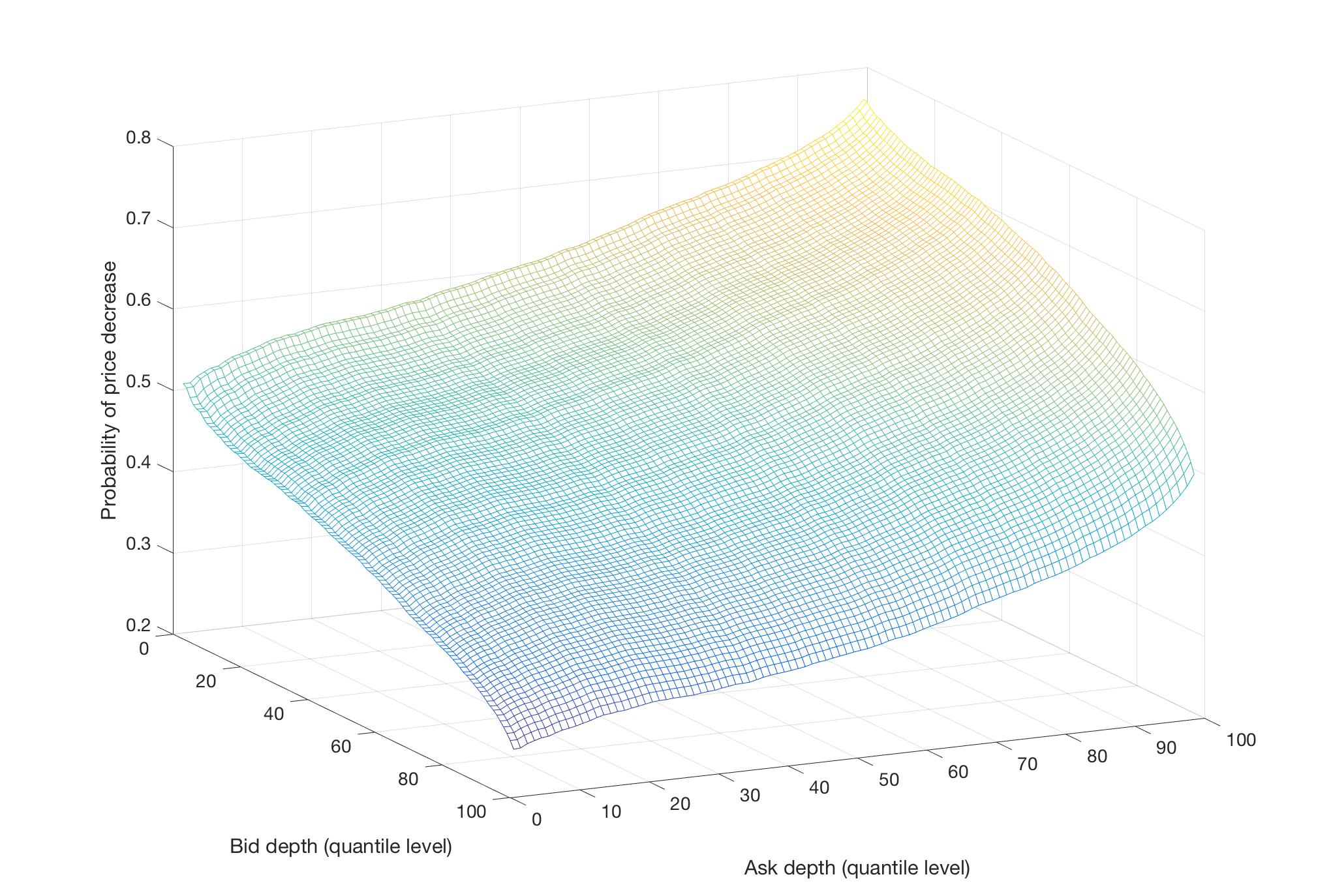}\
\includegraphics[width=.47\textwidth]{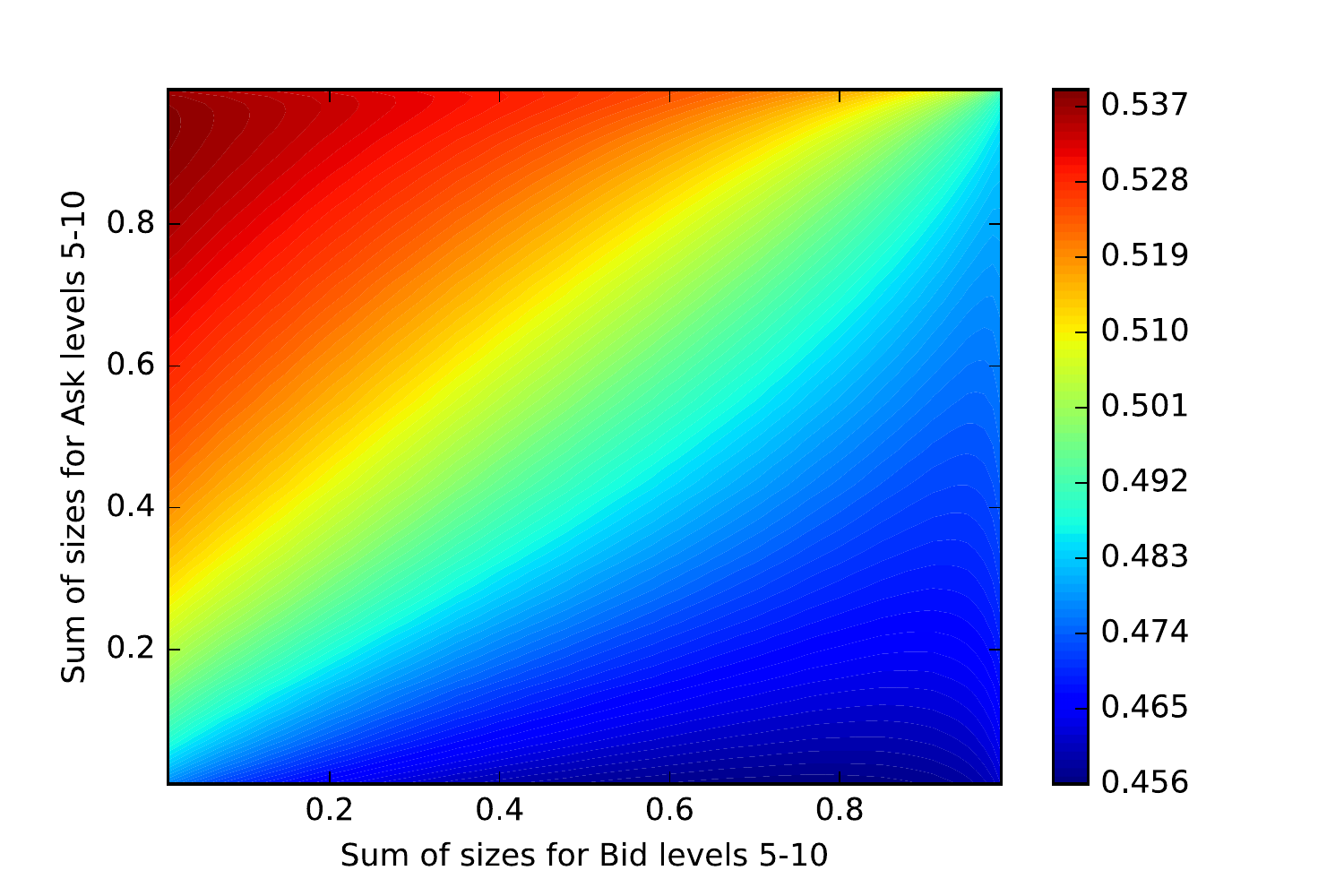}
\end{center}
\caption{Left: relation between depth at the bid, depth at the ask and the probability of a price decrease.
The $x$-axis and $y$-axis display the quantile level corresponding to the observed bid and ask depth.
Right: Contour plot displaying the influence of levels deeper in the order book (5 to 10) on the probability of a price decrease. }
\label{Contour1}
\end{figure}
Figure \ref{Contour1} presents an examples of such a feature: the relation between the depth on the bid and ask sides of the order book and the probability of a price decrease. 
Such relations have been studied in queueing models of limit order book dynamics \cite{Rama2,Rama3}. In particular, it was shown in \cite{Rama3} that when the order flow is symmetric then there exists a `universal' relation --- not dependent on model parameters --- between bid depth, ask depth and the probability of a price decrease at the next price move. However, the derivations in these models hinge on many statistical assumptions which may or may not hold, and the universality of such relations remained to be empirically verified.

Our analysis shows that there is indeed evidence for such a universal relation, across a wide range of assets and time periods. Figure  \ref{Contour1} (left) displays the probability of a price decrease as a function of the depth (the number of shares) at the best bid/ask price. 
The larger the best ask size, the more likely the  next price prove will be downwards. The probability is approximately constant along the center diagonal where the bid/ask imbalance is zero. However, as observed in queueing models \cite{Rama2,Rama3,Lopez1}, even under simplifying assumptions, the relation between this probability and various measures of the bid/ask imbalance is not linear. Furthermore,  such queueing models typically focus on the influence of depth at the top of the order book and it is more difficult to extract information from deeper levels of the order book. 
The right contour plot in Figure \ref{Contour1} displays the influence of limit orders deeper in the order book (here: total size aggregated across levels 5 to 10) on the probability of a price decrease. We see that the influence is less than the depth at the top of the book, as illustrated by the tighter range of predicted probabilities, but still significant.

\subsection{Universality across assets } \label{UniversalModelResults}
\begin{figure}[ht!]
\centering
\includegraphics[scale=0.6]{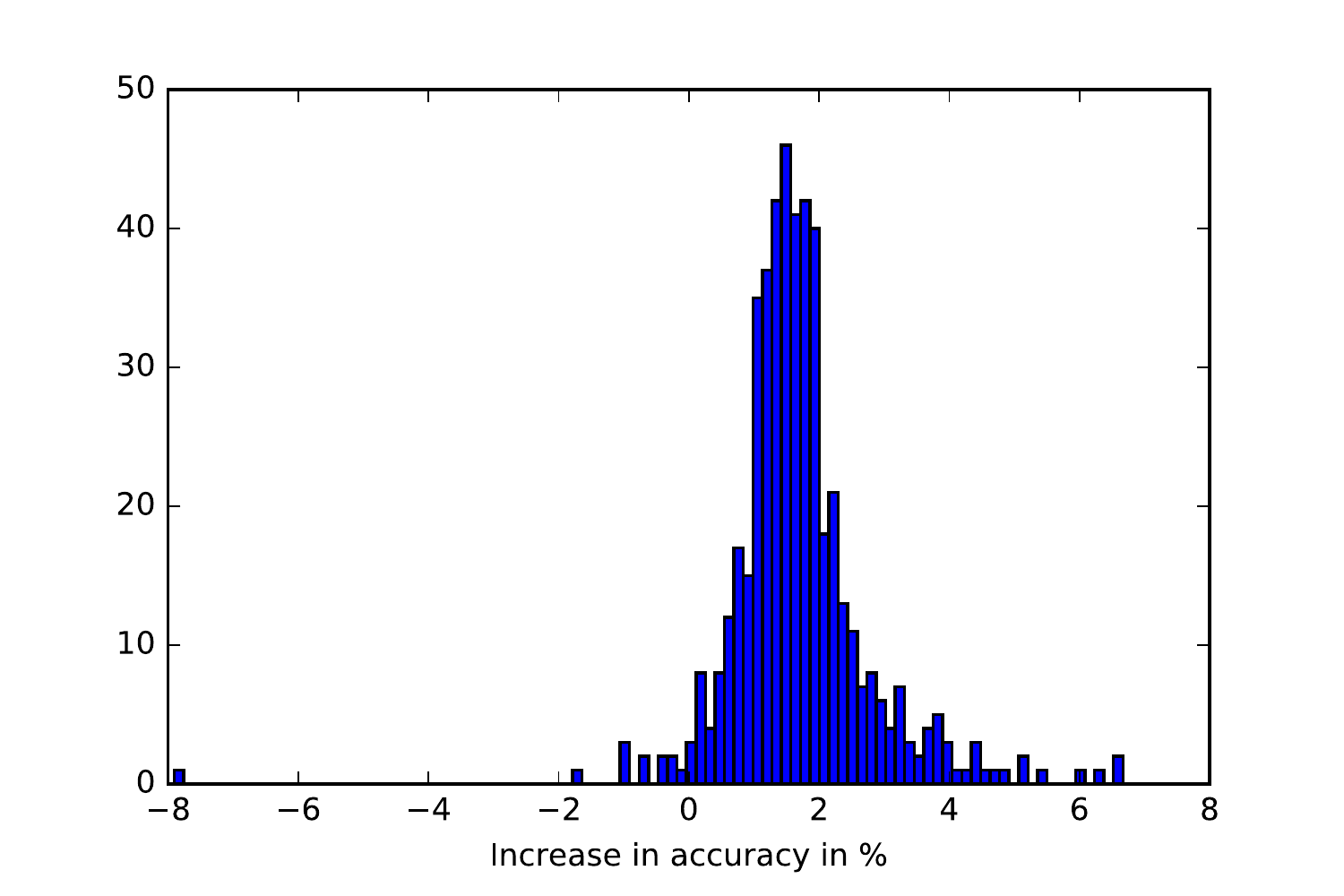}
\caption{Out-of-sample forecasting accuracy of the universal model compared with stock-specific models. Both are deep neural networks with $3$ LSTM layers followed by a ReLU layer. All layers have $50$ units. Models are trained to predict the direction of the next  move. Comparison across $489$ stocks, June-August, 2015. }
\label{URNN50_20vsRNN50_20WED}
\end{figure}

A striking aspect of our results is the stability across stocks of the features uncovered by the deep learning model, and its ability to extrapolate (`generalize') to stocks which it was not trained on. This may be illustrated by comparing 
forecasting accuracy of stock-specific models, trained only on data of a given stock, to  a universal model trained on a pooled data set of $500$ stocks, a much larger but extremely heterogeneous data set. As shown in Figure \ref{URNN50_20vsRNN50_20WED}, the universal model consistently outperforms the stock-specific models. This indicates there are common features, relevant to forecasting, across all stocks. Features extracted from data on stock A may be relevant to forecasting of price moves for stock B. 

Given the heterogeneity of the data, one might imagine that time series from different stocks should be first normalized (by average daily volume, average price or volatility etc.) before pooling them. Surprisingly, this appears {\it not} to be the case: we have observed that standard data transformations such as normalizations based on average volume, volatility or average spread, or partitioning the training data into sectors or categories such as large/small tick stocks do not improve training results. For example, a deep learning model trained on small tick stocks does not outperform the universal model in terms of forecasting price moves for small tick stocks. It appears that the model arrives at its own data-driven normalization of inputs based on statistical features of the data rather than ad hoc criteria.  

%Regarding intraday seasonality, the timestamp is included as part of the inputs so the model is able to learn adjustments for intraday seasonalities in a nonparametric way.

\begin{figure}[ht!]
\centering
\includegraphics[scale=0.6]{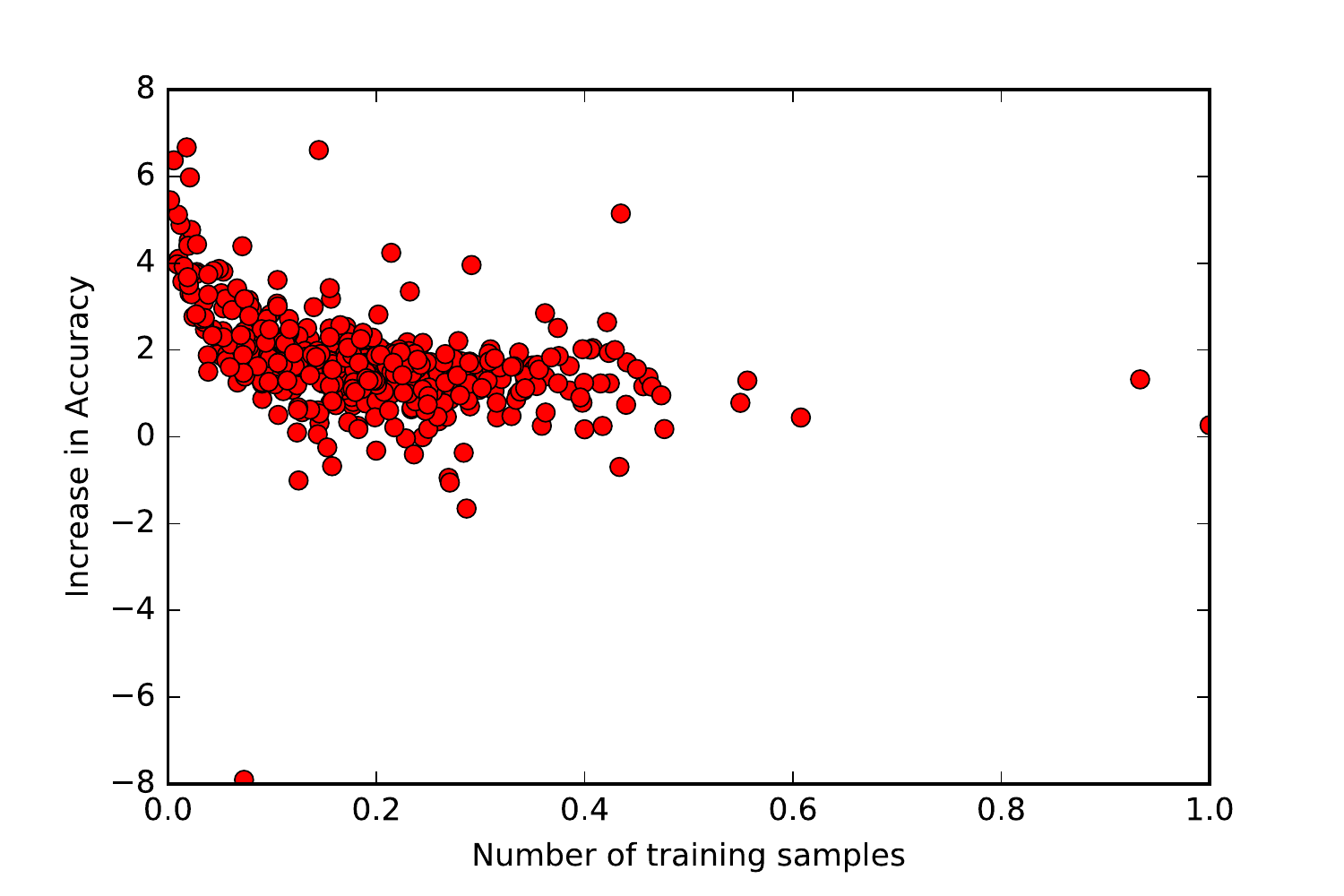}
\caption{Increase in out-of-sample forecast accuracy (in \%) of the universal model compared to stock-specific model, as a function of size of training set for stock specific model (normalized by total sample size, $N=24.1$ million). Models are trained to predict the direction  of next price move. Comparison across $500$ stocks, June-August, 2015. }
\label{PerformanceVersusDataLength}
\end{figure}

The source of the universal model's outperformance is well-demonstrated by Figure \ref{PerformanceVersusDataLength}. The universal model most strongly outperforms the stock-specific models on stocks with less data. The stock-specific model is more exposed to overfitting due to the smaller dataset while the universal model is able to \emph{generalize} by interpolating across the rich scenario space of the pooled data set and therefore is less exposed to overfitting.  
So, the existence of these common features seems to argue for pooling the data from different stocks, notwithstanding their heterogeneity, leading to a much richer and larger set of training scenarios.
Using 1 year of the pooled data set is roughly equivalent  to using 500 years (!) of data for training a single-stock model and the richness of the scenario space is actually {\it enhanced} by the diversity and heterogeneity of behavior across stocks.

\begin{figure}[ht!]
\centering
\includegraphics[scale=0.6]{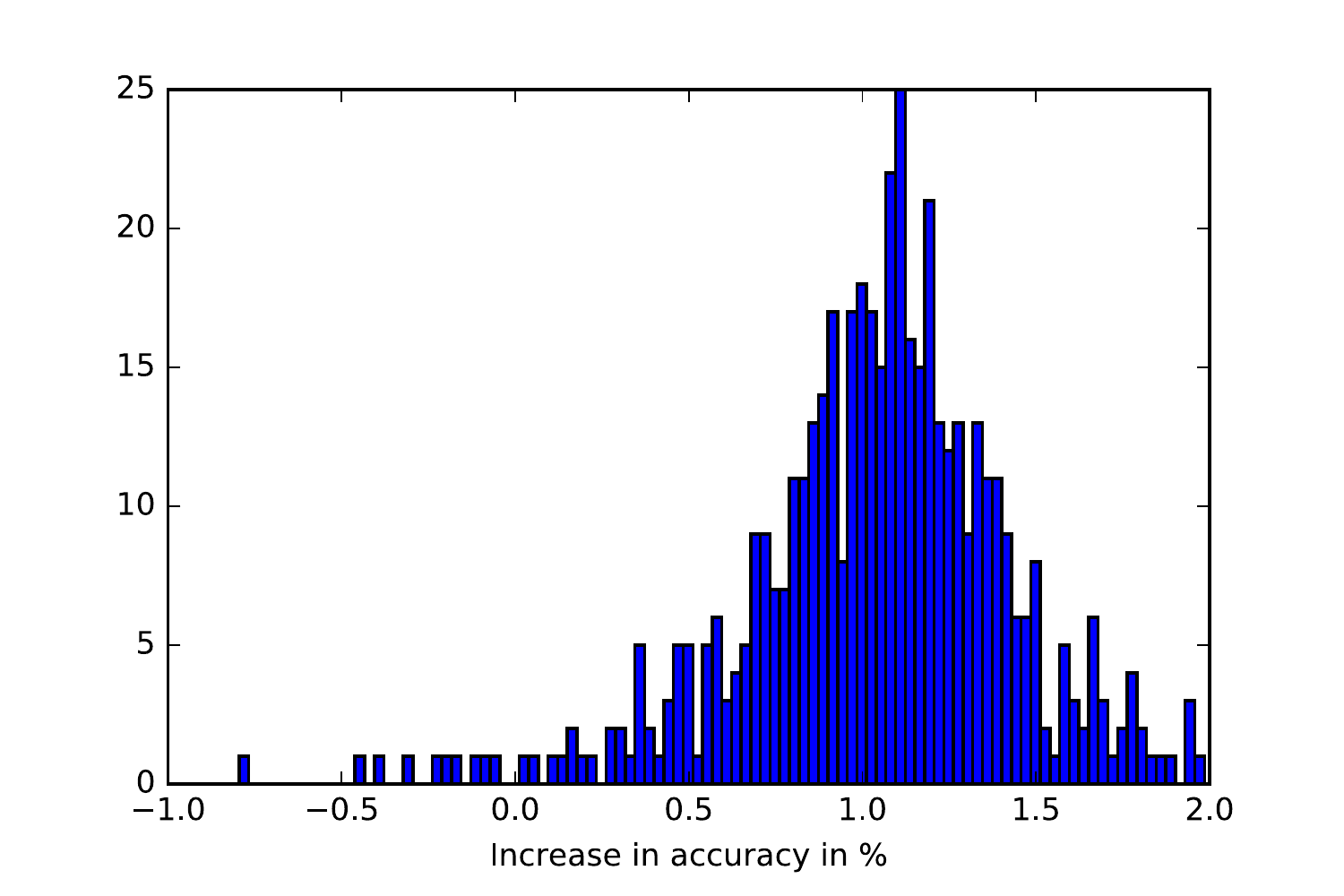}
\caption{Comparison of two universal models: a $150$ unit per layer model versus $50$ unit per layer model. Models are trained to predict direction $\{-1,+1\}$ of next mid-price move.  Out-of-sample prediction accuracy for direction of next price move, across approximately $500$ stocks (June-August, 2015).}
\label{150unit}
\end{figure}

Due to the large amount of data, very large universal models can be estimated without overfitting. Figure \ref{150unit} shows the increase in accuracy for a universal model with 150 units per layer (which amounts to several hundred thousand parameters) versus a universal model with 50 units per layer.

Remarkably, the universal model is even able to {\it generalize} to stocks which were not part of the training sample: if the model is only trained on data from stocks $\{1, \ldots, N \}$, its forecast accuracy is similar for stock $N+1$. This implies that the universal model is capturing features in the relation between order flow and price variastions which are common to all stocks. Table \ref{NewStocks} illustrates the forecast accuracy of  a universal model trained \emph{only} on stocks $1-464$ (for January 2014-May 2015), and tested on stocks $465-489$ (for June-August 2015). This universal model outperforms stock-specific models for stocks $465-489$, even though the universal model has never seen data from these stocks in the training set. The universal model trained only on stocks $1-464$ performs roughly the same for stocks $465-489$ as the universal model trained on the entire dataset of stocks $1-489$. Results are reported in Table \ref{NewStocks}.  

Figure \ref{500CompletelyNew} displays the accuracy of the universal model for $500$ completely new stocks, which are not part of the training sample. The universal model achieves a high accuracy on these new stocks, demonstrating that it is able to generalize to assets that are not included in the training data.
This is especially relevant for applications, where missing data issues, stock splits, new listings and corporate events constantly modify the universe of stocks.

\begin{table}[ht!]
\begin{center}
\scalebox{.9}{
 \begin{tabular}{  |c| c  | c| }
   \hline
Model  & Comparison  & Average  increase in accuracy   \\ \hline \hline
Stock-specific  &    25/25 & 1.45\%     \\ \hline
Universal &    4/25 &  -0.15\%    \\ \hline
\end{tabular} }
\end{center}
\caption{Comparison of universal model trained on stocks 1-464 versus (1) stock-specific models for stocks 465-489 and (2) universal model trained on all stocks 1-489. Models are trained to predict direction of next mid-price move. Second column shows the fraction of stocks where the universal model trained only on stocks 1-464 outperforms models (1) and (2).  The third column shows the average increase in accuracy. Comparison for $25$ stocks and out-of-sample results reported for June-August, 2015. }
\label{NewStocks}
\end{table}

\begin{figure}[h!]
\begin{center}
\includegraphics[width=.5\textwidth, height=60mm]{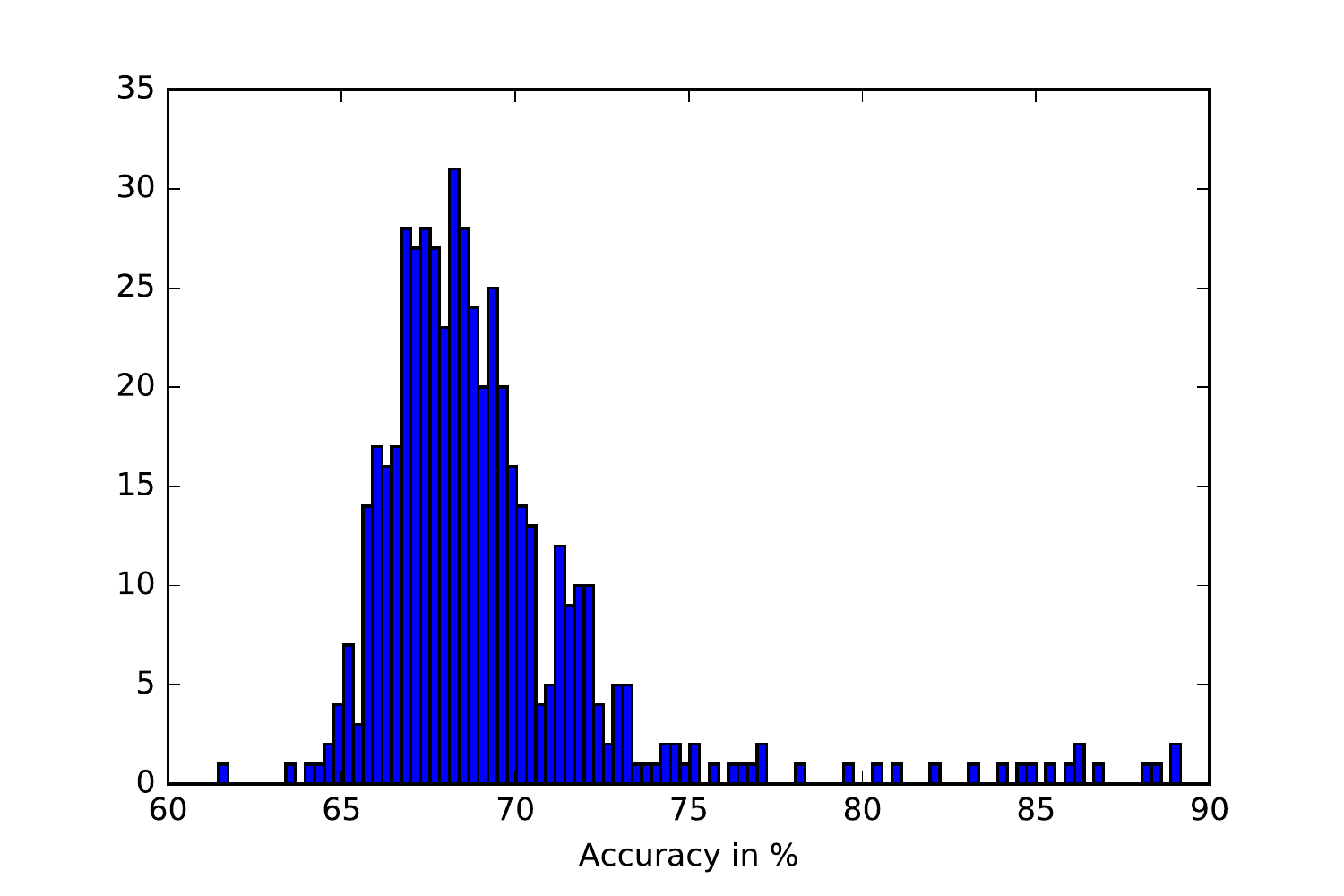}
\end{center}
\caption{Performance on approximately $500$ new stocks which the model has never seen before. Out-of-sample accuracy reported for June-August, 2015. Universal model trained during time period January 2014-May 2015.}
\label{500CompletelyNew}
\end{figure}

\subsection{Stationarity} \label{sec.stationarity}
The relationships uncovered by the deep learning model are not only stable across stocks but also stationary in time. This is illustrated by examining how forecast accuracy behaves when the training period and test period are separated in time.

Figure \ref{500CompletelyNew} shows the accuracy of the universal model on $500$  stocks which were not part of the training sample. The left histogram displays the accuracy in June-August, 2015, shortly after the training period (January 2014-May 2015), while the right plot displays the cross-sectional distribution of accuracy for the {\it same} model in January-March, 2017, {\it 18 months after the training period}. Interestingly, even one year after the training period, the forecasting accuracy is stable, without {\it any} adjustments. 

Such stability contrasts with the common practice of `recalibrating' models based on a moving window of recent data due to perceived non-stationarity. If the data were non-stationary, accuracy would decrease with the time span separating the training set and the prediction period and it would be better to train models only on recent periods immediately before the test set. However, we observe that this is not the case:
 Table \ref{NonStationarity} reports forecast results for models trained over periods extending up to $1, 3, 6,$ and 19 months before the test set. Model accuracy consistently increases as the length of the training set is increased.   The message is simple: use all available data, rather than an arbitrarily chosen time window.
 
 Note that these results are {\it not} incompatible with the data itself being non-stationary. The stability we refer to is the stability of the {\it relation} between the inputs (order flow and price history) and outputs (forecasts). If the inputs themselves are non-stationary, the output will be non-stationary but that does not contradict our point in any way.
\begin{figure}[h!]
\begin{center}
\includegraphics[width=.4\textwidth, height=50mm]{2015_500_never_trained_on}
\includegraphics[width=.4\textwidth, height=50mm]{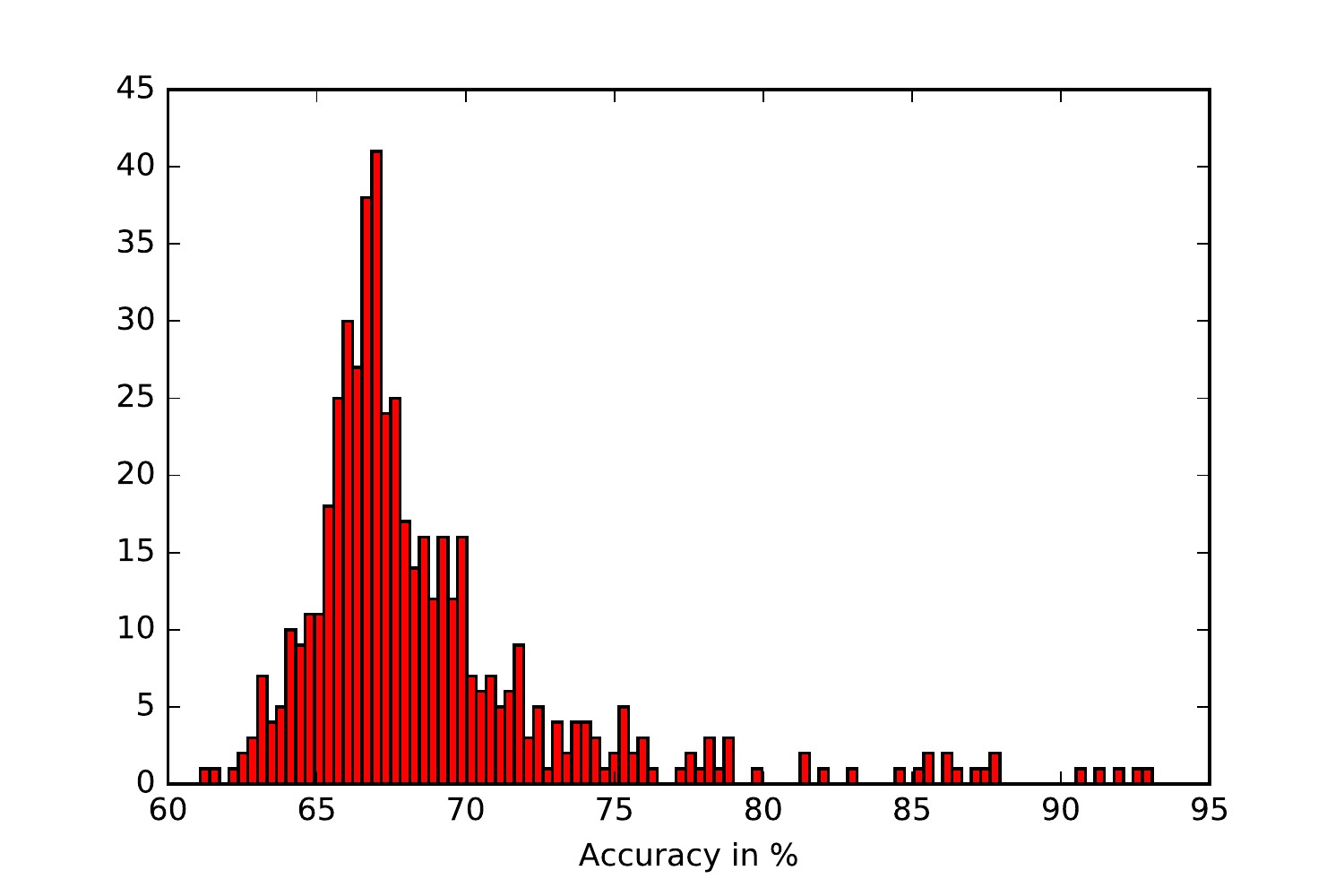}
\end{center}
\caption{Performance on  $500$ new stocks which the model has never seen before. \textcolor{blue}{Left}: out-of-sample accuracy reported for June-August, 2015. \textcolor{red}{Right}: out-of-sample accuracy reported for January-March, 2017. Universal model trained on data from January 2014-May 2015.}
\label{500CompletelyNew}
\end{figure}

\begin{table}[ht!]
\begin{center}
\scalebox{.9}{
 \begin{tabular}{  |c| c  | c| }
   \hline
Size of training set & \% of stocks for which 19-mo. training& Average increase in accuracy     \\ 
&  outperforms short-term training &for 19-month model \\
\hline %\hline
1 month   &    100\%  & 7.2\%     \\ \hline
3 months &    100\% & 3.7\%    \\ \hline
6 months &  100\%   & 1.6\%   \\ \hline
 \end{tabular} }
\end{center}
\caption{Out-of-sample forecast accuracy of deep learning models trained on entire training set (19 months) vs. deep learning models trained for shorter time periods immediately preceding the test period, across $50$ stocks Aug 2015. Models are trained to predict the direction of next price move. Second column shows the fraction of stocks where the $19$-th month model outperforms models trained on shorter time periods. The third column shows the average increase in accuracy across all stocks.  }
\label{NonStationarity}
\end{table}

\subsection{Path-dependence} \label{sec.temporal}

Statistical modeling of financial time series has been dominated by {\it Markovian } models which, for reasons of analytical tractability, assume that the evolution of the price and other state variables only depends on their current value and there is no added value to including their history beyond one lag. There is a trove of empirical evidence going against this hypothesis, and pointing to long-range dependence in financial time series \cite{Bacry,Lillo2004,Mandelbrot1997}. Our results are consistent with these findings: we find that the {\it history} of the limit order book contains significant additional information beyond that contained in its current state.

\begin{figure}[ht!]
\centering
\includegraphics[scale=0.6]{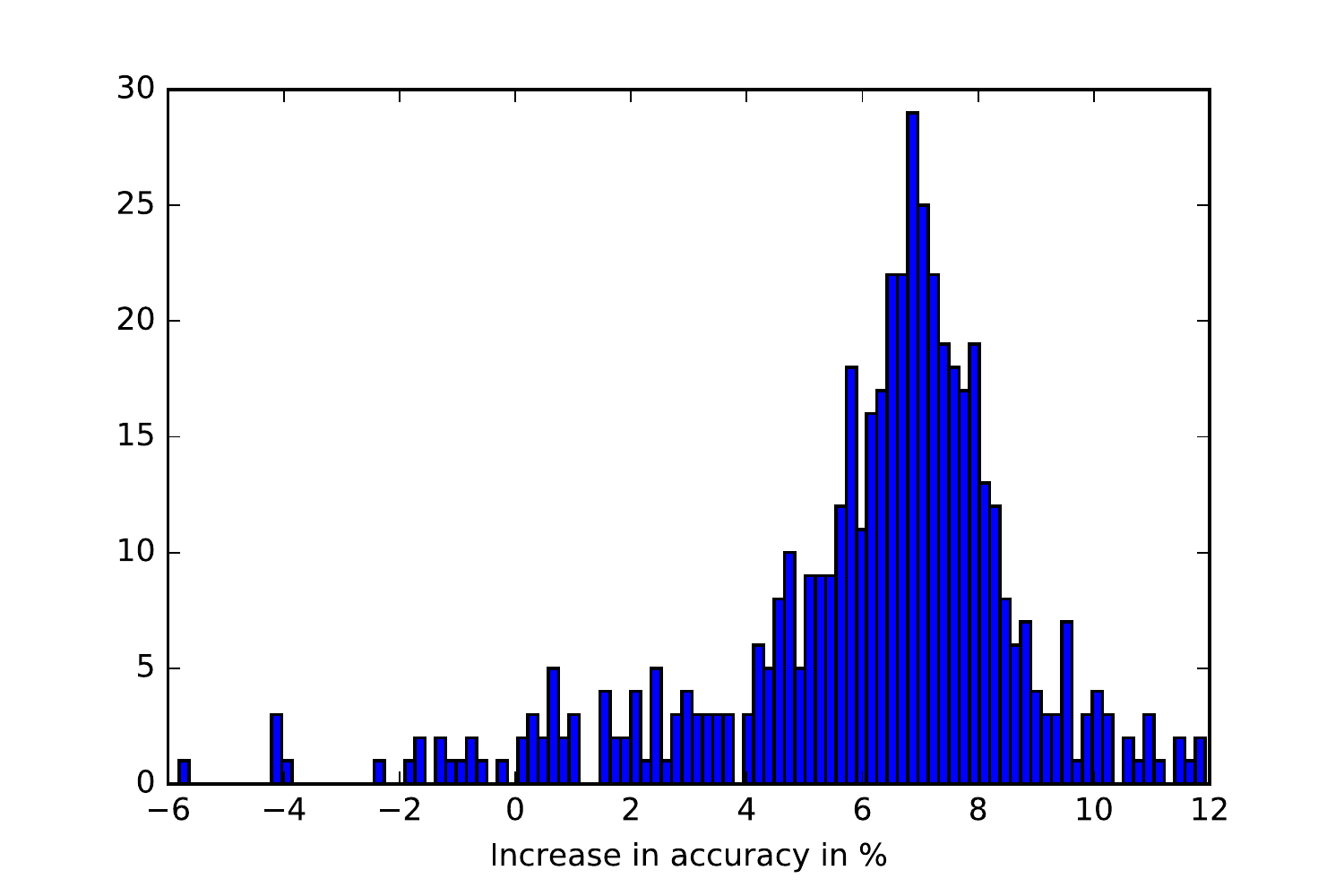}
\caption{Comparison of out-of-sample forecast accuracy of a LSTM network with a feedforward neural network trained to forecast the direction  of next  price move based on the current state of the limit order book. Cross-sectional results for $500$ stocks for test period June-August, 2015. }
\label{NNvsrnnHR2WED}
\end{figure}

Figure \ref{NNvsrnnHR2WED} shows the increase in accuracy when using an LSTM network, which is a function of the history of the order book, as  compared with a feedforward neural network, which is only a function of the most recent observation (a Markovian model).  The LSTM network, which incorporates temporal dependence, significantly outperforms the Markovian model. 

\begin{figure}[h!]
\begin{center}
\includegraphics[width=.6\textwidth, height=60mm]{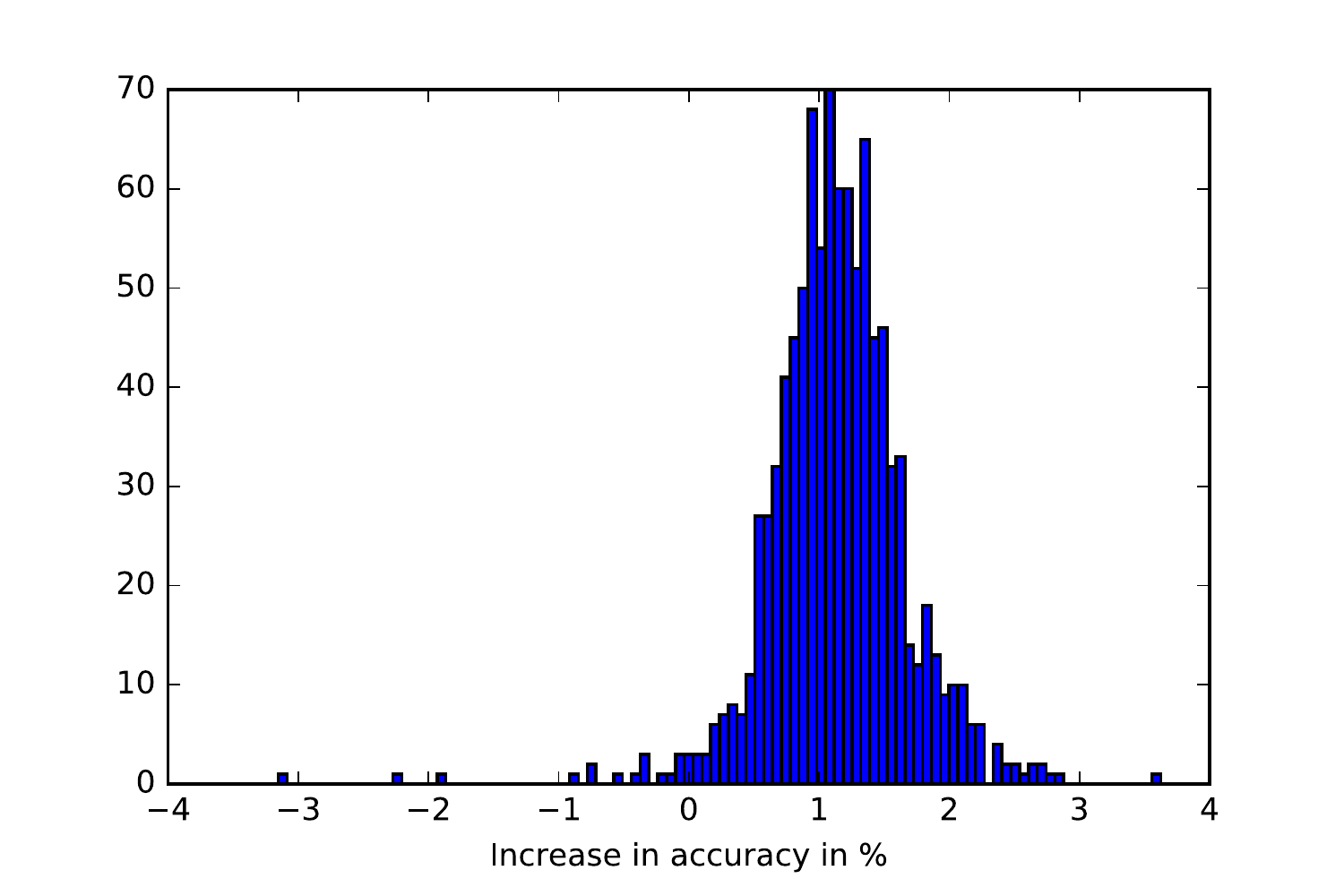}
\end{center}
\caption{Out-of-sample increase in accuracy when using \textcolor{red}{5000} lags versus \textcolor{blue}{100} lags, across $1,000$ stocks. Test  period : June-August 2015.}
\label{5000versus100steps}
\end{figure}

The accuracy of the forecast also increases when the network is provided with a longer history as input. 
Figure \ref{5000versus100steps} displays the accuracy of the LSTM network on a $5,000$-step sequence minus the accuracy of the LSTM network on a $100$-step sequence. Recall that a step $\Delta_{k} = \tau_{k+1} - \tau_k$ is on average $1.7$ seconds in the dataset so 5000 lags corresponds to 2 hours on average. There is a significant increase in accuracy, indicating that the deep learning model is able to find relationships between order flow and price change events over long time periods.  

Our results show that there is significant gain in model performance from including many lagged values of the observations in the input of the neural network, a signature of 
significant --- and exploitable --- temporal dependence in order book dynamics. 
%Sensitivity analyses can be used to analyze the model's dependence on specific events. Figure \ref{Yshift} shows the sensitivity to a perturbation in an input from $20$ time steps in the past. Specifically, we change the lagged price movement at time $t-20$ and observe the change in the model's prediction for time $t+1$.  

\section{Discussion}\label{sec.conclusion}

Using a  Deep Learning approach applied to a large dataset of billions of orders and transactions for 1000 US stocks, we have uncovered evidence of a {\bf universal price formation mechanism} relating history of the order book for a stock to the (next) price variation for that stock.
 More importantly, we are able to {\it learn} this mechanism through supervised training of a deep neural network on a high frequency time series of the limit order book. The resulting model displays several interesting features:
\begin{itemize}
\item Universality: the model is stable across stocks and sectors, and the model trained on all stocks outperforms  stock-specific models, even for stocks  not in  the training sample, showing that features captured are not stock-specific.
\item Stationarity:  model performance is stable across time,  even a year out of sample.
\item Evidence of `long memory' in price formation: including order flow history as input, even up to several hours, improves prediction performance.
\item Generalization: the model extrapolates well to  stocks not included in the training sample. This is especially useful since it demonstrates its applicability to recently listed instruments or those with incomplete or short data histories.
\end{itemize}
Our results illustrate the applicability and usefulness of Deep Learning methods for modeling of intraday behavior of financial markets. 
In addition to the fundamental insights they provide on the nature of price formation in financial markets, these findings have practical implications for model estimation and design. Training a single universal model is orders of magnitude less complex and costly than training  or estimating thousands of single-asset models. Since the universal model can generalize to new stocks (without training on their historical data), it can also be applied to newly issued stocks or stocks with shorter data histories.

\end{document}